\documentclass[11pt, twocolumn]{aastex6}

\newcommand{\lya}{\mbox{${\rm Ly}\alpha$}}

\newcommand{\HI}{\ion{H}{1}}

\usepackage{graphicx,amsmath,amssymb,epsfig}
\usepackage[T1]{fontenc}
\usepackage{multirow}
\usepackage{color}

\shorttitle{Identifying Radio Bright Quasars}
\shortauthors{Krogager et al.}

\begin{document}


\title{MALS--NOT: Identifying Radio-Bright Quasars\\ for the MeerKAT Absorption Line Survey}

\author{
J.-K. Krogager\altaffilmark{1,2},
N. Gupta\altaffilmark{3},
P. Noterdaeme\altaffilmark{1},
A. Ranjan\altaffilmark{1},
J. P. U. Fynbo\altaffilmark{4},\\
R. Srianand\altaffilmark{3},
P. Petitjean\altaffilmark{1},
F. Combes\altaffilmark{5},
A. Mahabal\altaffilmark{6}
}

\altaffiltext{1}{Institut d'Astrophysique de Paris, CNRS-UPMC, UMR7095, 98bis bd Arago, 75014 Paris, France}
\altaffiltext{2}{Dark Cosmology Centre, Niels Bohr Institute, University of
Copenhagen, Juliane Maries Vej 30, 2100 Copenhagen \O, Denmark}
\altaffiltext{3}{Inter-University Centre for Astronomy and Astrophysics, Post Bag 4, Ganeshkhind, 411 007 Pune, India}
\altaffiltext{4}{Niels Bohr Institute, University of
Copenhagen, Juliane Maries Vej 30, 2100 Copenhagen \O, Denmark}
\altaffiltext{5}{Observatoire de Paris, LERMA, CNRS: UMR8112, 61 Av. de l'Observatoire, 75014, Paris, France}
\altaffiltext{6}{California Institute of Technology, 1200 E. California Blvd, Pasadena, CA 91125, USA}

\begin{abstract}

\noindent
We present a preparatory spectroscopic survey to identify radio-bright, high-redshift quasars for the MeerKAT Absorption Line Survey (MALS). The candidates have been selected on the basis of a single flux density limit at 1.4~GHz ($>200$~mJy) together with mid-infrared color criteria from the {\it Wide-field Infrared Survey Explorer} ({\it WISE}). Through spectroscopic observations using the Nordic Optical Telescope, we identify 72 quasars out of 99 candidates targeted. We measure the spectroscopic redshifts based on characteristic, broad emission lines present in the spectra. Of these 72 quasars, 64 and 48 objects are at sufficiently high redshift ($z>0.6$ and $z>1.4$) to be used for the L-band and UHF-band spectroscopic follow-up with the Square Kilometre Array (SKA) precursor in South Africa: the MeerKAT.

\end{abstract}


\section{Introduction}
Extragalactic \ion{H}{1} 21-cm and OH absorption line surveys have been mostly limited to known samples of damped Ly$\alpha$ absorbers (DLAs) or \ion{Mg}{2} absorbers, from which about 50 detections of \HI\ absorption have been reported \citep{Briggs1983, Kanekar2003, Kanekar2014, Gupta2009, Curran2010, Gupta2012, Srianand2012, Dutta2017}. 
At $z>0.1$, only 5 OH molecular systems are known; in two of these cases absorbing gas is associated with the active galactic nucleus (AGN) itself, and in the remaining three cases absorption arises from the intervening lensing galaxy \citep[e.g.,][]{Darling2004, Kanekar2004}.
These scarce detections have led to inconclusive interpretations of the cold gas evolution in galaxies. In addition these data are biased by various preselections from optical surveys.

Thanks to large absorption line surveys with upcoming Square Kilometre Array (SKA) precursors and pathfinders, i.e., the Australia Square Kilometer Array Pathfinder \citep[][ASKAP]{Johnston2007, Allison2017}, APERTIF \citep{Apertif, Maccagni2017} and MeerKAT \citep{Gibbon2015, MALS, Jarvis2017}, the number of absorption line systems, especially sight lines through dusty ISM, at radio wavelengths is expected to dramatically increase over next few years.

The MeerKAT radio telescope in South Africa will be the most sensitive facility at 21-cm until the completion of SKA phase-I. Using this state-of-the-art telescope, the MeerKAT absorption line survey \citep[MALS,][]{MALS} will probe \ion{H}{1} 21-cm absorption out to redshift $z=1.5$ for about 1000 blind sightlines, i.e., no preselection based on optical absorption (Ly$\alpha$, \ion{Mg}{2}, etc.) is used.

The main goal of the survey is to quantify the redshift evolution of the cold gas fraction in and around galaxies out to redshift $z=1.5$ (and out to $z=1.8$ using OH absorption). The cold gas phase holds important clues for the regulation of the interstellar medium, and ultimately leads to the onset of star formation. At high redshift, the best way to study the cold neutral medium is through absorption studies \citep[e.g.][]{Srianand2005, Srianand2012, Ledoux2015, Noterdaeme2017}, since emission line studies only detect the brightest pockets of gas and therefore do not give a representative picture of the cold gas fraction through cosmic time.
Furthermore, studying the cold gas using 21-cm absorption on a purely radio-selected sample overcomes one of the main limitations of optically selected samples:
the possible dust-induced bias. The extent to which optically selected quasar samples are biased against dusty (and thereby preferentially cold and metal-rich) absorption systems has been discussed extensively through the years \citep{Pei1991, Ellison2001c, Jorgenson2006, Vladilo2005, Pontzen2009, Kaplan10, Fynbo2013a, Glikman2013, Krogager2015, Krogager2016b, Murphy2016}. Current data suggest that the average reddening from quasar absorbers is small, yet increases with metallicity, and it has been argued that the total cosmic metal density might be underestimated by up to a factor of two \citep{Pontzen2009}.
While the issue of a dust bias is more efficiently probed in radio selected samples, previous surveys have been too small to draw firm conclusions \citep{Ellison2005}. For this reason, the MALS sample will provide a unique opportunity to study cold gas absorbers in an unbiased way.

Moreover, the MALS sample will be useful for a range of auxiliary scientific goals; The large sample of \ion{H}{1} and OH absorbers obtained from the survey will (i) lead to tightest constraints on the fundamental constants of physics \citep{Kanekar2005, Rahmani2012}, and (ii) be ideally suited to probe the evolution of magnetic fields in disks of galaxies via Zeeman splitting or rotation measure synthesis \citep{Farnes2014}. The survey will also provide an unbiased census of \ion{H}{1} and OH absorbers, i.e., cold gas associated with powerful AGN ($>1024$~W~Hz$^{-1}$) at $0 < z < 2$ to investigate fundamental issues related to AGN evolution and feedback \citep{Maccagni2017}, and will simultaneously deliver a blind \ion{H}{1} and OH emission line survey, and radio continuum survey.
For more details about the individual science goals of MALS, we refer to the survey description \citep{MALS}.

The MALS survey design is specified in Table 2 of \citet{MALS}. Specifically, to maximize the optical depth sensitivity and redshift coverage, the L- and UHF-band pointings will be centred at flat-spectrum radio quasars (FSRQs) at $z>0.6$ and $z>1.4$, respectively. Bright FSRQs with known spectroscopic redshifts, especially in the southern hemisphere, are rare;
Therefore, we have initiated a large campaign to identify more high-redshift, radio-bright quasars using infrared color criteria to limit the influence of dust. The infrared-selected radio quasar candidates are then followed up using optical spectroscopy to classify the nature of the source and to measure its redshift. For the spectroscopic observations, higher priority is given to candidates that are likely to be FSRQs. Due to the large number of targets needed to be identified, the task has been split among several observing facilities, and in this paper, we report on one part of the survey carried out at the Nordic Optical Telescope (NOT) at the Observatorio del Roque de los Muchachos on La Palma, Spain, from which we can access the most northern targets of the overall survey. The results from the Southern African Large Telescope (SALT) campaign will be presented in a future paper. The MeerKAT observing campaign is expected to start in 2018 and the survey will be carried out over a period of 5 years.

The paper is structured as follows. In Section~\ref{selection}, we present the target selection criteria; in Section~\ref{observations}, we describe the optical follow-up and the data processing; in Section~\ref{results}, we present the results; and in Section~\ref{summary}, we provide a short discussion of caveats and a summary of our findings.

\section{Target Selection}
\label{selection}
First step was to assemble a catalog of southern ($\delta<+20$~deg) radio sources that are bright enough ($>$200~mJy) to meet the requirements of MALS in 1$-$2~hrs of telescope integration time. For this purpose, we used the NRAO VLA Sky Survey \citep[NVSS;][]{Condon1998} at 1.4 GHz and the Sydney University Molonglo Sky Survey \citep[SUMSS;][]{Bock1999, Mauch2003} at 843 MHz\footnote{A spectral index of $0.8$ was assumed to determine $1.4$ GHz flux densities of targets from SUMSS.}.
 The radio source catalog was then cross-correlated with photometry from the {\it Wide-field Infrared Survey Explorer} [{\it WISE}\,, \citealt{WISE}]\,). The {\it WISE} photometry covers 4 bands, here designated as $W_1$ (3.4~$\mu$m),  $W_2$ (4.6~$\mu$m), $W_3$ (12~$\mu$m), and $W_4$ (22~$\mu$m). The cross-matching is performed on the coordinates using a matching radius of 2~arcsec which corresponds to the astrometric uncertainty of the radio sources.
 The {\it WISE} color-space for radio sources identified as quasars from the Sloan Digital Sky Survey \citep[SDSS,][]{York2000} spectroscopic catalog is shown in Fig.~\ref{fig:wise}. The color-coding of each bin in the color-color diagram shows the average redshift of sources in the given bin. It is clearly seen that low-redshift quasars tend to cluster in one part of this color-space and we can therefore limit the number of low-redshift interlopers by applying the following photometric cuts \citep[similar to fig.~1 of][]{Krogager2016b}:
 $$ W_1 - W_2 < 1.3 \times (W_2 - W_3) -3.04 $$
	and $$ W_1 - W_2 > 0.6~.$$

The lower limit on the $W_1-W_2$ color is imposed to limit the number of stars and galaxies in our sample, since the contamination fraction increases rapidly when expanding to lower values of $W_1-W_2$ \citep[e.g.,][]{Stern2012, Richards2015}.
\citet{Stern2012} estimate the purity of quasars with $W_1-W_2>0.6$ to be $\sim$70\%.

For the NOT observations we limited the observations to northern targets, i.e., $-20$\degr $ < \delta < +20$\degr. This resulted in a parent sample of 196 targets.
Objects with compact radio morphology and flat spectral shapes are given priority as these are likely to be compact at cm-wavelengths and hence preferred for the intervening absorption studies \citep{Gupta2012}. In total we reject 41 targets with complex or extended radio morphologies based on available arcsec scale L-band images in FIRST \citep{FIRST}.

Furthermore, 14 sources are discarded based on extended optical morphology (classified as galaxy in SDSS) as these are dominated by low-redshift ($z<1.4$) sources\footnote{During the first night of observations we observed 8 sources with extended morphology and only 1 was a $z>1.4$ quasar, and the morphological classification of this source is most likely affected by a very bright nearby star.}. Five sources are discarded due to overlap with the plane of the Milky Way (closer than $\pm$15~deg Galactic latitude) to minimize the contamination from stars. We exclude two sources since they are bright stars based on the finding charts, three sources have been excluded since spectroscopic redshifts from SDSS are already available, and one source is excluded due to a catalog error, which was not discovered until after the observing runs.
The final sample of candidates is thus made up of 130 targets.\\

\section{Observations and Data Processing}
\label{observations}
During two observing runs in August 2016 (P53-012) and February 2017 (P54-005), we have observed a total of 99 candidate quasars with the (NOT) using the Andaluc\'ia faint object spectrograph and camera (ALFOSC). For all observations, we binned the CCD pixels by a factor of 2 along the wavelength axis and we aligned the slit with the parallactic angle. The candidates were observed using grism \#4 which covers the wavelength range from about 3200~\AA\ to 9100~\AA\ at a resolution of about $R \sim 300$ (depending on the slit-width used). During the observations we matched the slit-width according the seeing in order to limit slit-losses.
No blocking filter was used for the observations to maximize throughput. We thus caution, that the spectra suffer from second order contamination effects. However, as our main purpose of the spectroscopic follow-up was merely to secure the classification and to measure redshifts, the second order contamination is not of great importance.
The exposure time and slit-width used for the observation of each target is summarized in Table~\ref{tab:overview}. The seeing varied quite significantly in the Feb run ranging from 0.5 arcsec in the best case to $>$2~arcsec in the worst case. For the Aug run, the seeing was more stable around $\sim$1 arcsec. Photometric properties for the observed sample are summarized in Table~\ref{tab:phot} in Appendix~\ref{app:figures}.

During the observations, we gave priority to targets with a visible optical counterpart in the finding chart or sources that had optical data available, such as SDSS photometry. This allowed us to identify an optical counterpart in acquisition images of typically 10-20 seconds, reaching depths of $R \sim 21$~mag. In total, 40 targets had no visible counterpart in the finding charts. Since the finding charts were based on rather shallow photometry from the Digitized Sky Survey, we also attempted to look for optical counterparts where no obvious sources were visible in the finding charts by making deeper (1 minute, reaching $R\sim 22.5$~mag) acquisition images centred on the radio position. The results from these blind positions are summarized in Sect.~\ref{results}.\\

We used He and Ne calibration lamps for the wavelength calibration and on each night we observed at least one spectroscopic standard star in order to calibrate the response of the instrument. 

Firstly, the raw frames were bias subtracted and flat-field corrected using Halogen lamp flats. The background was then subtracted row by row (the spectra were dispersed vertically on the CCD) using a low order Chebyshev polynomial fit with robust sigma-clipping.
Cosmic rays were rejected using the method developed by \citet{vanDokkum2001}. We use the Python implementation, {\tt astroscrappy}\footnote{Written by Curtis McCully, see documentation on GitHub: \url{https://github.com/astropy/astroscrappy}}, which is significantly faster than the original implementation in IRAF\footnote{IRAF is distributed by the National Optical Astronomy Observatory, which is operated by the Association of Universities for Research in Astronomy (AURA) under cooperative agreement with the National Science Foundation.}.

Extraction and calibration of the 1-dimensional (1D) spectra was performed using the IRAF tasks {\tt apall}, {\tt identify}, {\tt sensitivity}, and {\tt calibrate}. The respective vacuum wavelength solutions and sensitivity functions were applied to the 2D spectral frames in order to obtain calibrated 2D spectra.\\

\begin{deluxetable*}{lccccccl}
\tabletypesize{\small}	
\tablecaption{Overview of spectroscopic sample \label{tab:overview}}
\tablehead{
\colhead{Target}          &     \colhead{R.A.}      &     \colhead{Decl.}  & \colhead{F$_{1.4 {\rm GHz}}$} & \colhead{Exp. time}& \colhead{Slit width} & \colhead{$z_{\rm spec}$} & \colhead{Classification} \\
                &               &                 & \colhead{(mJy)}  &  \colhead{(sec)}  & \colhead{(arcsec)}  &                &    \\
}
\startdata
MALS0017$-$1256   &  00:17:08.03  &  $-$12:56:24.9  &    929.7   &    300  &      1.0  &   0.870  &  Quasar  \\
MALS0022$+$0608   &  00:22:32.46  &  $+$06:08:04.6  &    339.7   &    120  &      1.3  &  --      &  Blazar?  \\
MALS0042$+$1246   &  00:42:43.06  &  $+$12:46:57.6  &    635.0   &    360  &      1.3  &   2.150  &  Quasar  \\
MALS0051$+$1747   &  00:51:47.15  &  $+$17:47:10.3  &   1367.5   &    480  &      1.3  &   0      &  Galaxy  \\
MALS0053$-$0702   &  00:53:15.65  &  $-$07:02:33.4  &    248.2   &    450  &      1.0  &   2.130  &  Quasar  \\
MALS0105$+$1845   &  01:05:49.69  &  $+$18:45:07.1  &    339.5   &    600  &      1.0  &  --      &  Blazar?  \\
MALS0117$-$0425   &  01:17:27.84  &  $-$04:25:11.5  &    421.4   &   1200  &      1.0  &   1.720  &  Quasar  \\
MALS0211$+$1707   &  02:11:48.77  &  $+$17:07:23.2  &    539.8   &    600  &      1.3  &   1.980  &  Quasar  \\
MALS0216$+$1724   &  02:16:50.70  &  $+$17:24:04.9  &    395.2   &    300  &      1.3  &   1.530  &  Quasar  \\
MALS0226$+$0937   &  02:26:13.72  &  $+$09:37:26.3  &    374.6   &     60  &      1.3  &   2.610  &  Quasar  \\
MALS0226$+$1941   &  02:26:39.92  &  $+$19:41:10.1  &    209.8   &    900  &      1.0  &   2.190  &  Quasar  \\
MALS0240$+$0832   &  02:40:46.80  &  $+$08:32:58.3  &    305.4   &    300  &      1.3  &   1.100  &  Quasar  \\
MALS0249$+$0440   &  02:49:39.93  &  $+$04:40:28.9  &    420.5   &    420  &      1.3  &   2.010  &  Quasar  \\
MALS0249$+$1237   &  02:49:44.50  &  $+$12:37:06.3  &    261.2   &    300  &      1.3  &   0      &  Star/galaxy  \\
MALS0256$-$0218   &  02:56:30.36  &  $-$02:18:44.5  &    260.3   &    300  &      1.0  &   1.970  &  Quasar  \\
MALS0304$-$1126   &  03:04:13.80  &  $-$11:26:53.5  &    335.0   &   1200  &      1.0  &   1.550  &  Quasar  \\
MALS0314$-$0909   &  03:14:07.94  &  $-$09:09:46.9  &    226.8   &   1200  &      1.0  &   0.312  &  Emission line galaxy  \\
MALS0317$+$0655   &  03:17:06.76  &  $+$06:55:50.3  &    211.8   &   1800  &      1.3  &   0.275  &  Quasar  \\
MALS0328$-$0152   &  03:28:08.59  &  $-$01:52:20.2  &    221.9   &    600  &      1.0  &   2.681  &  Quasar  \\
MALS0341$+$1519   &  03:41:57.75  &  $+$15:19:25.5  &    200.4   &   2700  &      1.0  &   1.699  &  Quasar, blind observation  \\
MALS0350$-$0351   &  03:50:16.86  &  $-$03:51:11.3  &    230.5   &    600  &      1.0  &   0.611  &  Quasar  \\
MALS0356$+$1900   &  03:56:33.46  &  $+$19:00:34.6  &   1051.3   &    420  &      1.0  &   1.480  &  Quasar  \\
MALS0356$-$0831   &  03:56:34.55  &  $-$08:31:21.3  &    239.7   &   1800  &      1.0  &   1.447  &  Quasar  \\
MALS0455$+$1850   &  04:55:41.91  &  $+$18:50:10.9  &    328.1   &   1500  &      1.0  &   0.548  &  Quasar  \\
MALS0510$-$1959   &  05:10:24.23  &  $-$19:59:50.3  &    336.2   &   1200  &      1.0  &   0.472  &  Quasar  \\
MALS0512$+$1517   &  05:12:40.99  &  $+$15:17:23.8  &    966.5   &    600  &      1.0  &   2.568  &  Quasar  \\
MALS0516$+$0732   &  05:16:56.35  &  $+$07:32:52.7  &    231.7   &    600  &      1.0  &   2.594  &  Quasar  \\
MALS0519$+$1746   &  05:19:38.34  &  $+$17:46:41.2  &    319.1   &   1200  &      1.0  &   0  &  Galaxy, early type  \\
MALS0529$-$1126   &  05:29:05.55  &  $-$11:26:07.5  &    455.8   &    900  &      1.0  &   0.995  &  Quasar  \\
MALS0600$-$0710   &  06:00:12.92  &  $-$07:10:37.9  &    224.9   &   1800  &      1.0  &   0  &  Galaxy/Star  \\
MALS0730$+$1739   &  07:30:37.09  &  $+$17:39:51.5  &    694.2   &   3600  &      1.0  &  --      &  N/A, blind observation  \\
MALS0731$+$1433   &  07:31:59.01  &  $+$14:33:36.3  &    316.5   &    300  &      1.0  &   2.632  &  Quasar  \\
MALS0824$-$1029   &  08:24:44.81  &  $-$10:29:43.6  &    345.4   &    300  &      1.0  &   0  &  Post-starburst Galaxy  \\
MALS0836$+$0406   &  08:36:01.35  &  $+$04:06:36.0  &    435.3   &   3600  &      1.0  &   0.743  &  Galaxy, late type, blind observation  \\
MALS0909$-$1637   &  09:09:10.66  &  $-$16:37:53.8  &    340.1   &   1800  &      1.0  &   2.471  &  Quasar  \\
MALS0909$-$0500   &  09:09:16.99  &  $-$05:00:53.2  &    457.2   &   1500  &      1.0  &   0.403  &  Galaxy, late type  \\
MALS0910$-$0526   &  09:10:51.01  &  $-$05:26:26.8  &    337.9   &    300  &      1.0  &   2.387  &  Quasar  \\
MALS0939$-$1731   &  09:39:19.21  &  $-$17:31:35.4  &    263.3   &    600  &      1.0  &   1.831  &  Quasar  \\
MALS0941$-$0321   &  09:41:46.01  &  $-$03:21:21.7  &    346.2   &   1800  &      1.0  &   1.849  &  Quasar  \\
MALS0951$-$0601   &  09:51:38.34  &  $-$06:01:38.0  &    378.8   &    600  &      1.0  &   1.839  &  Quasar  \\
MALS1007$-$1817   &  10:07:43.51  &  $-$18:17:32.4  &    205.2   &    900  &      1.0  &   0.493  &  Galaxy, late type  \\
MALS1008$-$0959   &  10:08:02.27  &  $-$09:59:19.3  &    645.3   &    600  &      1.0  &   1.688  &  Quasar  \\
MALS1009$+$1602   &  10:09:54.53  &  $+$16:02:03.8  &    201.9   &    900  &      1.0  &   1.770: &  Quasar - redshift tentative  \\
MALS1024$-$0852   &  10:24:44.61  &  $-$08:52:06.4  &    463.8   &   3600  &      1.3  &   0.468  &  Emission line galaxy, blind observation  \\
MALS1033$-$1210   &  10:33:35.61  &  $-$12:10:32.1  &   2065.1   &    900  &      1.0  &   1.328  &  Emission line galaxy  \\
MALS1043$+$1641   &  10:43:47.34  &  $+$16:41:06.9  &    259.1   &   1800  &      1.0  &   0.399  &  Galaxy, late type  \\
MALS1046$+$1535   &  10:46:08.12  &  $+$15:35:36.6  &    430.9   &   3600  &      1.0  &   --     &  N/A, blind observation  \\
MALS1050$-$0318   &  10:50:24.43  &  $-$03:18:08.8  &    243.5   &    420  &      1.0  &   1.806  &  Quasar  \\
MALS1059$-$1118   &  10:59:55.36  &  $-$11:18:18.7  &    314.5   &    900  &      1.0  &   1.942  &  Quasar  \\
MALS1119$-$0527   &  11:19:17.36  &  $-$05:27:07.9  &   1174.4   &    600  &      1.0  &   2.651  &  Quasar  \\
MALS1124$-$1501   &  11:24:02.56  &  $-$15:01:59.1  &    261.9   &    600  &      1.0  &   2.551  &  Quasar  \\
MALS1148$+$1404   &  11:48:25.45  &  $+$14:04:49.7  &    322.0   &   4000  &      1.0  &   0.727  &  Emission line galaxy, blind observation  \\
MALS1150$+$1317   &  11:50:43.70  &  $+$13:17:36.1  &    254.4   &    600  &      1.0  &   1.530  &  Quasar  \\
MALS1206$-$0714   &  12:06:32.23  &  $-$07:14:52.6  &    698.8   &   2400  &      1.3  &   2.263  &  Quasar  \\
MALS1215$-$0628   &  12:15:14.42  &  $-$06:28:03.5  &    360.4   &   1800  &      1.8  &   3.223  &  Quasar  \\
MALS1218$-$0631   &  12:18:36.18  &  $-$06:31:15.8  &    409.1   &   1200  &      1.8  &   0.658  &  Quasar  \\
MALS1219$-$1809   &  12:19:05.45  &  $-$18:09:11.2  &    300.3   &    600  &      1.0  &   0.189  &  Quasar  \\
MALS1221$-$1237   &  12:21:23.49  &  $-$12:37:24.1  &    343.7   &   1800  &      1.8  &   1.882  &  Quasar  \\
MALS1225$-$1144   &  12:25:24.48  &  $-$11:44:31.2  &    479.5   &   2700  &      1.0  &  --      &  N/A, blind observation  \\
MALS1231$-$1236   &  12:31:50.30  &  $-$12:36:37.5  &    276.0   &   1200  &      1.8  &   2.100  &  Quasar  \\
MALS1318$-$1441   &  13:18:56.01  &  $-$14:41:55.0  &    283.0   &   2000  &      1.8  &   0.632  &  Quasar  \\
MALS1343$-$0834   &  13:43:09.26  &  $-$08:34:57.5  &    241.5   &    720  &      1.0  &   1.077  &  Quasar  \\
MALS1351$-$1019   &  13:51:31.98  &  $-$10:19:32.9  &    726.1   &   1500  &      1.8  &   2.998  &  Quasar  \\
MALS1456$+$0456   &  14:56:25.83  &  $+$04:56:45.2  &    287.9   &   1200  &      1.0  &   2.136  &  Quasar  \\
MALS1623$+$1239   &  16:23:03.03  &  $+$12:39:58.4  &    523.3   &    300  &      1.0  &   0  &  Star/Galaxy  \\
MALS1625$-$0416   &  16:25:39.34  &  $-$04:16:16.3  &    257.2   &   1200  &      1.0  &   0  &  Galaxy, early type  \\
MALS1639$+$1144   &  16:39:06.46  &  $+$11:44:09.2  &    341.9   &    900  &      1.0  &   0  &  Star/Galaxy  \\
MALS1645$-$0432   &  16:45:52.10  &  $-$04:32:53.3  &    423.3   &   1800  &      1.0  &  --      &  N/A  \\
MALS1649$+$0626   &  16:49:50.51  &  $+$06:26:53.3  &    389.2   &    300  &      1.0  &   2.130  &  Quasar  \\
MALS1650$-$1248   &  16:50:38.03  &  $-$12:48:54.5  &    275.5   &    720  &      1.0  &   2.490  &  Quasar  \\
MALS1653$-$0102   &  16:53:57.67  &  $-$01:02:14.2  &    317.4   &    600  &      1.0  &   1.070  &  Quasar  \\
MALS1716$-$0613   &  17:16:27.09  &  $-$06:13:56.5  &    229.9   &    480  &      1.0  &   1.840  &  Quasar  \\
MALS1721$+$1626   &  17:21:05.79  &  $+$16:26:49.1  &    507.9   &    900  &      1.0  &   1.814  &  Quasar  \\
MALS1722$+$1652   &  17:22:39.45  &  $+$16:52:08.7  &    257.1   &    600  &      1.0  &  --      &  N/A  \\
MALS1724$+$0326   &  17:24:23.07  &  $+$03:26:32.5  &    217.8   &    450  &      1.3  &   0  &  Galaxy  \\
MALS1725$+$0622   &  17:25:07.45  &  $+$06:22:42.1  &    213.4   &    180  &      1.0  &   1.330  &  Quasar  \\
MALS2034$-$0523   &  20:34:25.65  &  $-$05:23:32.2  &    419.7   &    900  &      1.3  &   2.070  &  Quasar  \\
MALS2054$-$0932   &  20:54:56.08  &  $-$09:32:40.8  &    313.6   &    600  &      1.0  &   1.350  &  Quasar  \\
MALS2059$-$1440   &  20:59:59.61  &  $-$14:40:43.1  &   1018.1   &   1500  &      1.3  &   0.690  &  Galaxy  \\
MALS2120$+$1327   &  21:20:42.48  &  $+$13:27:24.2  &    401.5   &    600  &      1.3  &   1.110  &  Quasar  \\
MALS2139$+$1718   &  21:39:37.03  &  $+$17:18:26.9  &    219.4   &    360  &      1.3  &   1.700  &  Quasar  \\
MALS2158$+$0925   &  21:58:00.88  &  $+$09:25:46.4  &    437.7   &    600  &      1.3  &   0.445  &  Quasar  \\
MALS2201$+$0312   &  22:01:27.50  &  $+$03:12:15.6  &    300.5   &    900  &      1.3  &   2.190  &  Quasar  \\
MALS2220$+$1307   &  22:20:04.97  &  $+$13:07:12.1  &    812.0   &   1800  &      1.3  &   0.760  &  Quasar  \\
MALS2223$+$1213   &  22:23:59.11  &  $+$12:13:38.9  &    246.4   &   1080  &      1.0  &   0  &  Galaxy  \\
MALS2224$+$1304   &  22:24:20.03  &  $+$13:04:50.2  &    294.5   &   1800  &      1.3  &   0.685  &  Quasar  \\
MALS2230$+$0348   &  22:30:50.19  &  $+$03:48:36.8  &    242.8   &   3600  &      1.3  &   1.330  &  Quasar, blind observation  \\
MALS2238$-$1344   &  22:38:26.48  &  $-$13:44:22.6  &    241.2   &    600  &      1.3  &   0.257  &  Quasar  \\
MALS2243$+$1814   &  22:43:54.81  &  $+$18:14:45.9  &   1004.4   &    480  &      1.0  &   0.760  &  Quasar  \\
MALS2247$+$1211   &  22:47:05.52  &  $+$12:11:51.4  &    223.7   &   1800  &      1.0  &   2.185  &  Quasar  \\
MALS2300$+$1940   &  23:00:36.41  &  $+$19:40:02.9  &    210.4   &    360  &      1.3  &   2.160  &  Quasar  \\
MALS2300$+$0337   &  23:00:40.87  &  $+$03:37:10.3  &    509.0   &    600  &      1.3  &   1.860  &  Quasar  \\
MALS2308$-$1149   &  23:08:05.25  &  $-$11:49:45.7  &    256.8   &   1500  &      1.0  &   1.940  &  Quasar  \\
MALS2310$+$1114   &  23:10:02.89  &  $+$11:14:03.6  &    248.1   &    360  &      1.3  &   1.550  &  Quasar  \\
MALS2316$+$0429   &  23:16:34.61  &  $+$04:29:40.2  &    214.0   &    480  &      1.0  &   2.160  &  Quasar  \\
MALS2332$-$1423   &  23:32:31.61  &  $-$14:23:18.8  &    234.1   &   1500  &      1.0  &   0.524  &  Quasar  \\
MALS2338$-$1218   &  23:38:08.04  &  $-$12:18:51.4  &    450.9   &    600  &      1.0  &   1.185  &  Quasar  \\
MALS2340$+$0959   &  23:40:07.26  &  $+$09:59:59.0  &    205.0   &   4500  &      1.3  &   2.210  &  Quasar, blind observation  \\
MALS2359$+$1924   &  23:59:14.02  &  $+$19:24:20.6  &    284.1   &   1800  &      1.0  &   0.515  &  Quasar  \\[1mm]
\hline
MALS0845$-$1118  &  08:45:48.34  &  $-$11:18:00.1   &    264.4   &    --   &       --  &     --   &  No source in acquisition image.  \\
MALS0954$+$1252  &  09:54:14.75  &  $+$12:52:22.6   &    355.7   &    --   &       --  &     --   &  No source in acquisition image.  \\
MALS1130$-$1006  &  11:30:15.25  &  $-$10:06:38.4   &    215.2   &    --   &       --  &     --   &  No source in acquisition image.  \\
MALS2258$-$0958  &  22:58:13.47  &  $-$09:58:17.2   &    225.7   &    --   &       --  &     --   &  No source in acquisition image.  \\
\enddata
\end{deluxetable*}


\section{Results}
\label{results}
\begin{figure}
\includegraphics[width=0.98\columnwidth]{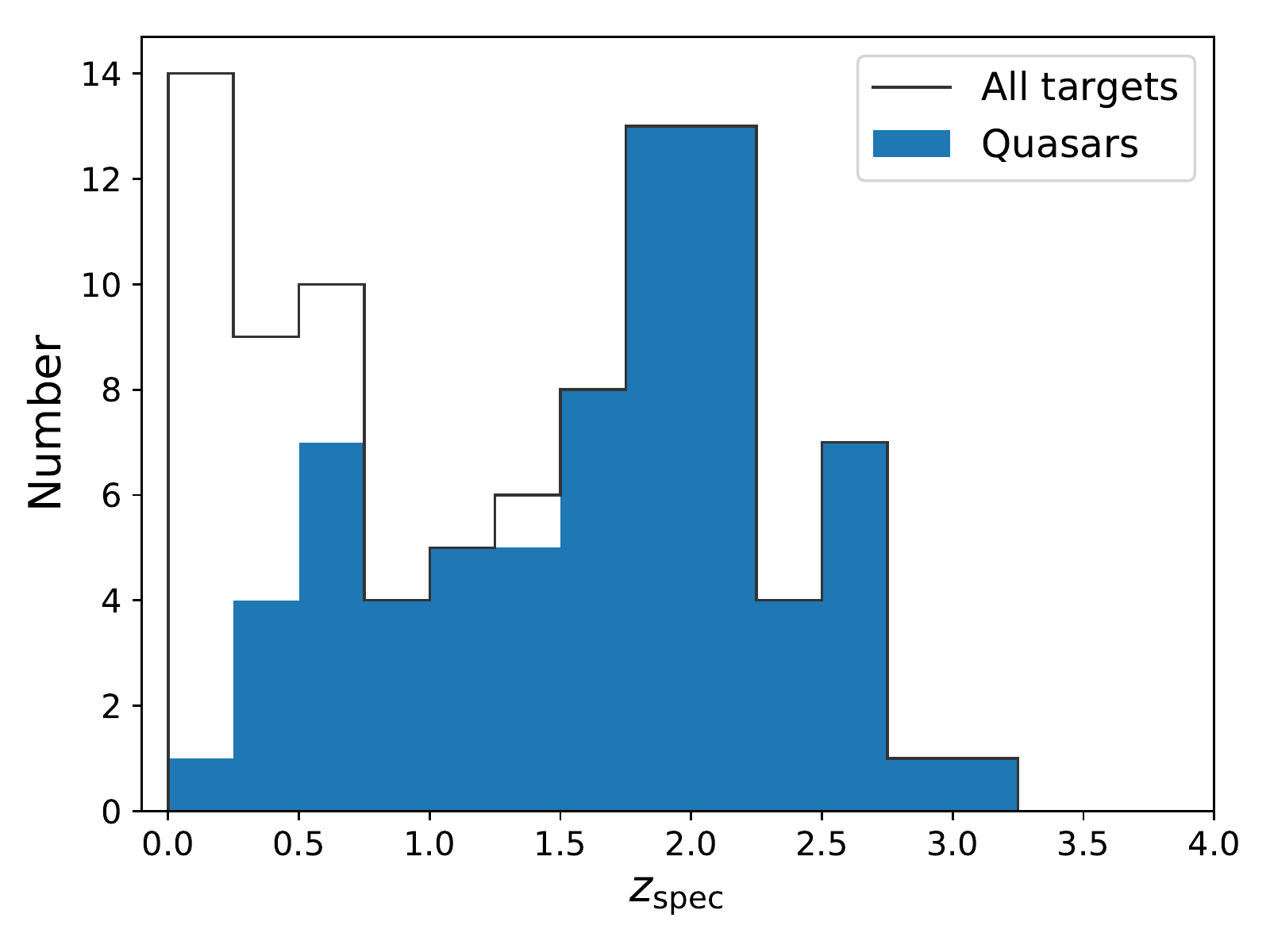}
\caption{Redshift distribution of spectroscopically observed targets.}\label{fig:zdist}
\end{figure}

\begin{figure*}
\centering
\includegraphics[width=2.\columnwidth]{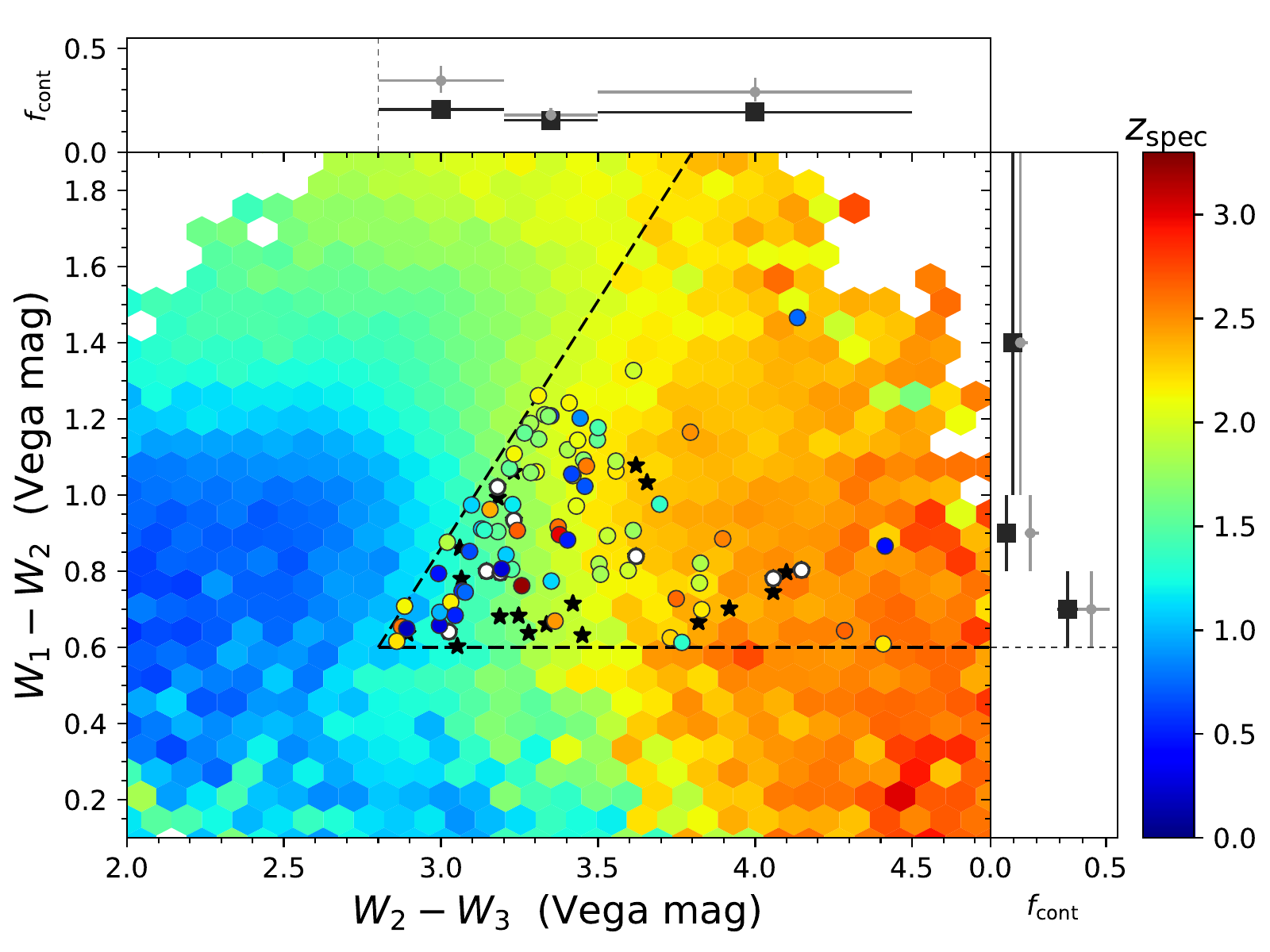}
\caption{{\it WISE} color-color plot for quasars. The underlying distribution shows spectroscopically observed quasars from the Sloan Digital Sky Survey \citep{Paris2017}. The color coding corresponds to the mean redshift of quasars in a given bin.
	The MALS quasars are shown as large, colored dots (following the same color coding). Black stars indicates contaminating stars or galaxies and white dots mark the candidates for which no classification was possible.
	The dashed lines show the region of color-space that we used to select high-redshift candidates. The two panels at the top and right edges indicate the contaminating fraction, $f_{\rm cont}$, of stars and galaxies in dark gray, and the contamination including the unidentified sources is shown in light gray.}\label{fig:wise}
\end{figure*}

We classify each observed target based on spectral features in the extracted 1D spectra. In most cases quasars are easily identified due to their broad emission lines, which furthermore enable us to measure its redshift.
The summary of all classifications and the measured redshifts are given in Table~\ref{tab:overview}. The uncertainty on the spectroscopic redshift is of the order $\sigma_z = 0.005$.
In cases where there are only narrow emission lines, we classify the target as an emission line galaxy and provide the measured redshift. In 8 cases we were not able to obtain a secure classification. These are marked either as unknown identification (`N/A'), blazar or tentative (redshift followed by `:') in Table~\ref{tab:overview}.
In total, we securely identify 72 quasars out of 99 observed targets, and 48 of these are at a sufficiently high redshift ($z>1.4$) for the UHF band spectroscopic observations of the MALS survey. For the L-band observations of MALS 64 out of the 72 quasars are at sufficiently high redshifts ($z>0.6$). In Fig.~\ref{fig:zdist}, we show the distribution of spectroscopic redshifts for all objects where it was possible to determine the redshift.
All targets are shown in the {\it WISE} color-color plot in Fig.~\ref{fig:wise}. The distribution of redshifts in the mid-infrared color-space follows largely the same pattern as the underlying SDSS sample taking into account the large scatter in the redshifts ($\pm0.8$) for each bin in color-space. In accordance with previous studies of quasars in the near-infrared color-space \citep[e.g.,][]{Assef2010, Stern2012}, we find that the fraction of contaminants decreases for increasing $W_1-W_2$ color.
A preview of the spectra are shown in Figs.~\ref{spec1d} and \ref{spec2d} in Appendix \ref{app:figures} (the full sets of figures are available online).

As mentioned in Sect.~\ref{observations}, we attempted to identify optical counterparts for 13 targets where no source was visible in the shallow finding chart. After a long acquisition image (typically 1 min), we were able to identify a faint source at the position of the radio source in 9 out of 13 cases. Out of these 9 blind observations, we were able to identify three quasars and three galaxies; three spectra did not result in any identification. These 9 blind observations are marked as `blind observation' in the classification notes of Table~\ref{tab:overview}. In the last 4 out of the 13 blind fields, we did not observe any optical counterpart in the deeper acquisition images and hence did not proceed with the spectroscopic observations. These are listed in the bottom part of Table~\ref{tab:overview} for completeness, but they are not counted as actual spectroscopic observations in this optical follow-up study. In order not to bias our survey, these unidentified radio sources will be part of our full spectroscopic campaign to search for \ion{H}{1} and OH absorption at radio wavelengths.\\

We searched the spectra for absorption lines either from the quasar host or from intervening systems. The objects with clearly visible absorption features are given in Table~\ref{tab:absorbers}. In cases where we were able to identify the lines, we also give the absorber redshift. We note that due to the low resolution and limited signal-to-noise ratio in the continuum, the identifications are tentative in most cases.

\begin{deluxetable}{ccr}
\tabletypesize{\small}	
\tablecaption{Absorption Systems \label{tab:absorbers}}
\tablehead{
\colhead{Target}       &    \colhead{$z_{\rm abs}$} & \colhead{Notes}
}
\startdata
MALS0053$-$0702   & \nodata &  No ID  \\
MALS0117$-$0425   &  1.315  &  \ion{Mg}{2}  \\
MALS0226$+$0937   & \nodata &  No ID  \\
MALS0226$+$1941   &  2.036  &  \ion{Mg}{2}  \\
MALS0256$-$0218   &  1.970  &  BAL  \\
MALS0328$-$0152   & \nodata &  No ID  \\
MALS0512$+$1517   &  2.568  &  Associated absorber  \\
MALS0516$+$0732   &  2.594  &  Associated absorber  \\
MALS0939$-$1731   &  1.319  &  \ion{Mg}{2}  \\
      ---         &  0.739  &  \ion{Mg}{2}  \\
MALS1351$-$1019   &  2.758  &  \lya\  \\
MALS1649$+$0626   &  2.310  &  Associated absorber  \\
MALS1716$-$0613   &  1.661  &  \ion{Mg}{2}  \\
MALS2139$+$1718   &  1.326  &  \ion{Mg}{2}  \\
MALS2247$+$1211   &  2.185  &  Associated absorber  \\
MALS2340$+$0959   & \nodata &  No ID  \\
\enddata
\end{deluxetable}

\section{Discussion and Summary}
\label{summary}
Using strong, broad emission lines as the only indicator for quasar identification is a simple and robust way to verify that the source is indeed a quasar and simultaneously allows a determination of the redshift. However, this approach comes with one main drawback: Not all quasars exhibit these broad, characteristic emission lines \citep{Meusinger2012, Meusinger2016, Krogager2015, Krogager2016b}. For the main science case of the MALS project, namely the study of intervening cold gas absorption, such a bias against non-standard quasars (incl. type-II and blazars) will have no significant implications since the bias is intrinsic to the quasars and not caused by the intervening absorbers. However, for the science case related to AGN fuelling and intrinsic properties of the AGN, the bias must be kept in mind.

From our observations at the NOT we can estimate the severeness of the bias from the number of unidentified spectra in our survey. In total, we found 8 targets with no clear identification and 72 quasars. As a conservative estimate, assuming that all unidentified sources are indeed atypical quasars without broad emission lines, we thus find a maximal missing fraction of 10\%.

The fraction of quasars in this region of {\it WISE} color space has independently been estimated at around 70\% \citep{Stern2012}, which agrees well with the observed number of quasars (72 out of 99). This indicates that the missing number of quasars is indeed small.

We note that some targets are classified as tentative blazars or atypical quasars, but for these targets we lack spectral features that allow us to determine the redshift. Thus, for the purpose of MALS, these targets are of little interest.\\

In summary, we have carried out a spectroscopic survey for radio loud quasars at high redshift. We used a single flux density limit at 1.4~GHz (requiring that candidates have $F_{1.4\rm GHz}>200$~mJy) together with simple color criteria in the $W_1-W_2$ and $W_2-W_3$ color-space to select candidates at $z>1.5$. Using the NOT we have carried out a survey of 99 candidates with compact radio morphologies. Out of the total number of candidates observed, 72 are quasars of which 64 and 48 objects are at sufficiently high redshift for the L- and UHF-band spectroscopic observations as part of MALS.

\acknowledgements

{\small
JKK acknowledges financial support from the Danish Council for Independent Research (EU-FP7 under the Marie-Curie grant agreement no. 600207) with reference DFF-MOBILEX--5051-00115.
AR, NG, RS, PN and PPJ acknowledge the support from Indo-French Centre for the Promotion of Advance Research (IFCPAR) under project number 5504-2.
Based on observations made with the Nordic Optical Telescope, operated on the island of La Palma jointly by Denmark, Finland, Iceland, Norway, and Sweden, in the Spanish Observatorio del Roque de los Muchachos of the Instituto de Astrof\'isica de Canarias.
The Digitized Sky Survey is based on photographic data obtained using The UK Schmidt Telescope. The UK Schmidt Telescope was operated by the Royal Observatory Edinburgh, with funding from the UK Science and Engineering Research Council, until 1988 June, and thereafter by the Anglo-Australian Observatory. Original plate material is copyright © the Royal Observatory Edinburgh and the Anglo-Australian Observatory. The plates were processed into the present compressed digital form with their permission. The Digitized Sky Survey was produced at the Space Telescope Science Institute under US Government grant NAG W-2166.
}

\newpage

\bibliographystyle{apj}


\appendix

\section{Supporting Figures and Tables}
\label{app:figures}

\begin{figure}
\includegraphics[width=0.95\columnwidth]{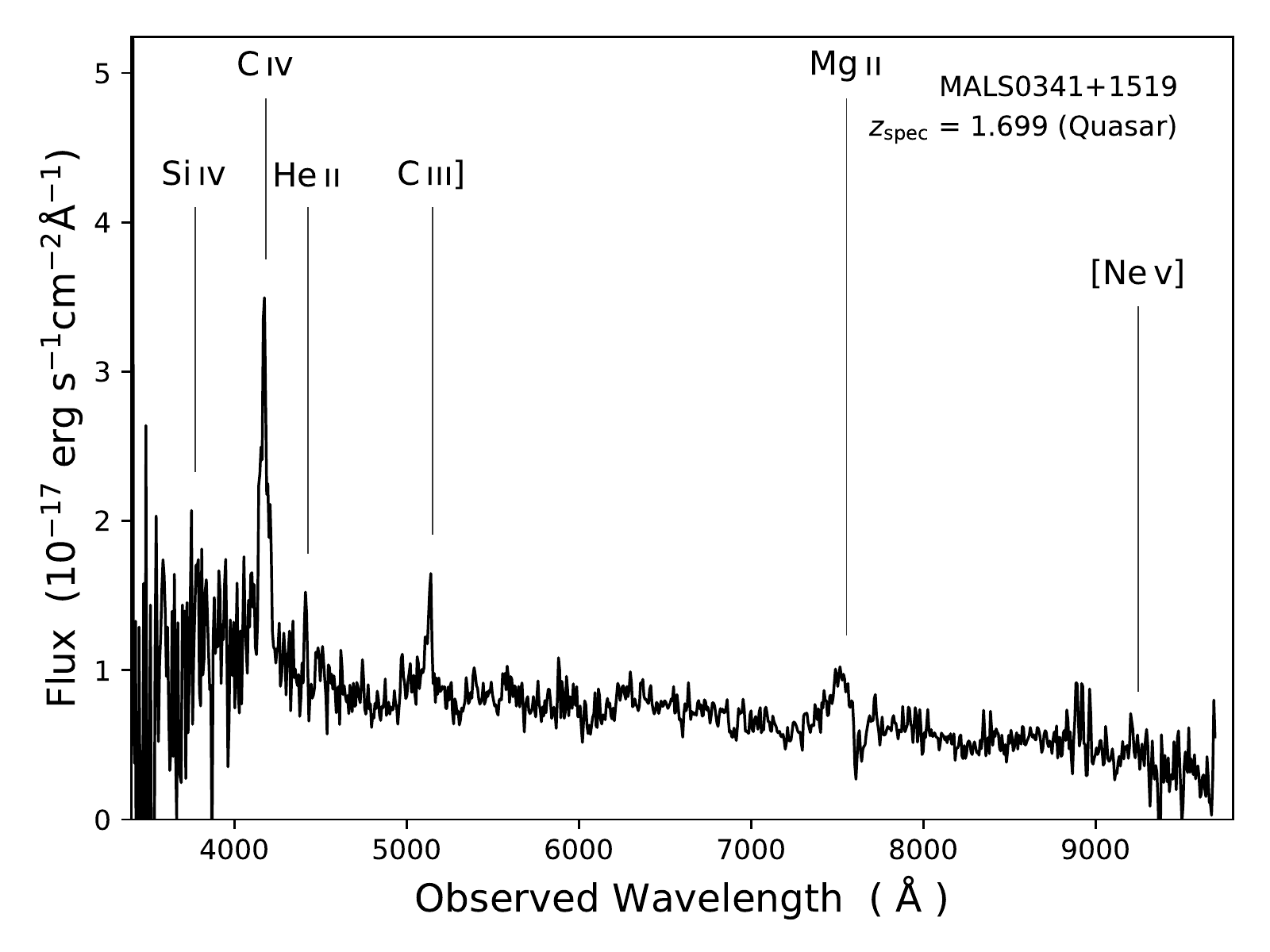}
\caption{Extracted 1-dimensional spectrum from the NOT. The location of various emission lines typically present in quasar spectra are shown for the redshift given in the upper right corner. A redshift of $z_{\rm spec}=-1$ means that no spectroscopic identification was possible, and a redshift of $z_{\rm spec}=0$ means that the target is classified as a star or galaxy with no emission features.\\
(The complete figure set (99 images) is available in the online journal and in the source files.)}\label{spec1d}
\end{figure}

\begin{figure}
\includegraphics[width=0.95\columnwidth]{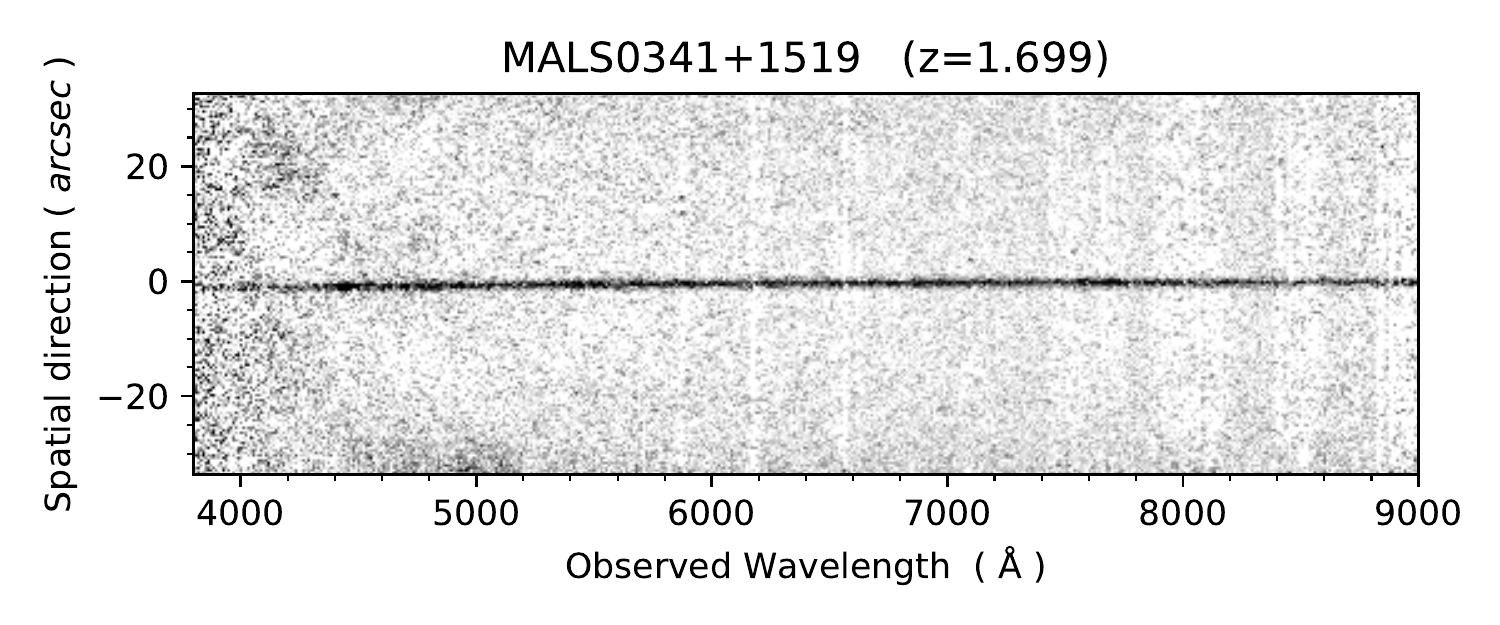}
\caption{Processed 2-dimensional spectrum.
	(The complete figure set (99 images) is available in the online journal and in the source files.)}\label{spec2d}
\end{figure}

For each 1D spectrum in Fig.~\ref{spec1d}, we show the extracted data in black and the associated emission lines are marked if a spectroscopic redshift has been determined (as indicated in the upper right corner). If not, then we give $z=-1$ in the upper right corner.
The 2D spectra in Fig.~\ref{spec2d} are shown with a linear color-scale matched to the noise level in each spectrum; specifically we use the median absolute deviation (MAD) to estimate the background noise and set the lower scale to $-1\times${\sc mad} and the upper scale to $+6\times${\sc mad}. Above each 2D spectrum we give the target name and the observing period (P53-012 or P54-005) in which the spectrum was obtained.

In Table~\ref{tab:phot}, we summarize the photometric properties of the spectroscopic sample compiling the radio brightness at 1.4~GHz, optical $g$ and $r$ band magnitudes from SDSS together with our rough estimate of the $r$-band magnitude from the observed spectrum. We note that the synthetic $r$-band magnitudes have not been corrected for possible slit losses and their typical uncertainties are of the order of $0.2$~mag. Lastly, we include mid-infrared photometry and colors from {\it WISE}.

\begin{deluxetable*}{lccccccccccl}
\tabletypesize{\small}
\tablecaption{Photometric properties of spectroscopic sample \label{tab:phot}}
\tablehead{
\colhead{Target} & \colhead{R.A.} & \colhead{Decl.} & \colhead{F$_{1.4~{\rm GHz}}$} & \colhead{$g$}& \colhead{$r$} & \colhead{$r'_{\rm syn}$} & \colhead{$W_1$}  & \colhead{$W_2$} & \colhead{$W_3$} & \colhead{$W_1 - W_2$} & \colhead{$W_2 - W_3$} \\
\colhead{(1)} & \colhead{(2)} & \colhead{(3)} & \colhead{(4)} & \colhead{(5)} & \colhead{(6)} & \colhead{(7)} & \colhead{(8)}  & \colhead{(9)} & \colhead{(10)} & \colhead{(11)} & \colhead{(12)}
}
\startdata
MALS0017$-$1256 & 00:17:08.03 & $-$12:56:24.9 &  929.7 &  20.47  &  19.72  & 20.0 & 14.16 & 12.96 & 9.51 & 1.20 & 3.44  \\ 
MALS0022+0608 & 00:22:32.46 & +06:08:04.6 &  339.7 &  18.99  &  18.54  & 18.5 & 13.23 & 12.21 & 9.03 & 1.02 & 3.18  \\ 
MALS0042+1246 & 00:42:43.06 & +12:46:57.6 &  635.0 &  20.15  &  19.93  & 20.4 & 15.78 & 14.54 & 11.13 & 1.24 & 3.41  \\ 
MALS0051+1747 & 00:51:47.15 & +17:47:10.3 & 1367.5 &  21.09  &  20.18  & 20.5 & 15.59 & 14.65 & 11.42 & 0.93 & 3.23  \\ 
MALS0053$-$0702 & 00:53:15.65 & $-$07:02:33.4 &  248.2 &  20.89  &  20.22  & 20.3 & 16.31 & 15.25 & 11.95 & 1.06 & 3.30  \\ 
MALS0105+1845 & 01:05:49.69 & +18:45:07.1 &  339.5 &  20.32  &  19.86  & 19.4 & 15.61 & 14.81 & 11.66 & 0.80 & 3.15  \\ 
MALS0117$-$0425 & 01:17:27.84 & $-$04:25:11.5 &  421.4 &  20.91  &  20.71  & 20.2 & 16.05 & 15.00 & 11.71 & 1.06 & 3.29  \\ 
MALS0211+1707 & 02:11:48.77 & +17:07:23.2 &  539.8 &  20.64  &  20.76  & 20.4 & 16.22 & 15.42 & 11.82 & 0.80 & 3.60  \\ 
MALS0216+1724 & 02:16:50.70 & +17:24:04.9 &  395.2 &  20.75  &  20.29  & 19.9 & 15.99 & 15.09 & 11.91 & 0.90 & 3.18  \\ 
MALS0226+0937 & 02:26:13.72 & +09:37:26.3 &  374.6 & \nodata & \nodata & 18.4 & 15.03 & 14.11 & 10.74 & 0.92 & 3.37  \\ 
MALS0226+1941 & 02:26:39.92 & +19:41:10.1 &  209.8 & \nodata & \nodata & 21.0 & 16.52 & 15.46 & 11.90 & 1.06 & 3.56  \\ 
MALS0240+0832 & 02:40:46.80 & +08:32:58.3 &  305.4 & \nodata & \nodata & 19.6 & 15.57 & 14.59 & 11.49 & 0.97 & 3.10  \\ 
MALS0249+0440 & 02:49:39.93 & +04:40:28.9 &  420.5 &  20.55  &  20.18  & 20.6 & 16.15 & 15.10 & 11.68 & 1.05 & 3.42  \\ 
MALS0249+1237 & 02:49:44.50 & +12:37:06.3 &  261.2 & \nodata & \nodata & 17.5 & 15.03 & 14.35 & 11.16 & 0.68 & 3.19  \\ 
MALS0256$-$0218 & 02:56:30.36 & $-$02:18:44.5 &  260.3 &  19.17  &  18.79  & 18.8 & 15.21 & 13.88 & 10.27 & 1.33 & 3.61  \\ 
MALS0304$-$1126 & 03:04:13.80 & $-$11:26:53.5 &  335.0 & \nodata & \nodata & 20.3 & 16.98 & 15.84 & 12.34 & 1.14 & 3.50  \\ 
MALS0314$-$0909 & 03:14:07.94 & $-$09:09:46.9 &  226.8 & \nodata & \nodata & 20.4 & 15.25 & 14.61 & 11.72 & 0.64 & 2.89  \\ 
MALS0317+0655 & 03:17:06.76 & +06:55:50.3 &  211.8 &  22.29  &  21.90  & 20.6 & 14.78 & 14.12 & 11.13 & 0.66 & 2.99  \\ 
MALS0328$-$0152 & 03:28:08.59 & $-$01:52:20.2 &  221.9 & \nodata & \nodata & 18.5 & 15.46 & 14.55 & 11.31 & 0.91 & 3.24  \\ 
MALS0341+1519 & 03:41:57.75 & +15:19:25.5 &  200.4 &  21.40  &  21.02  & 21.2 & 16.32 & 15.22 & 11.77 & 1.09 & 3.45  \\ 
MALS0350$-$0351 & 03:50:16.86 & $-$03:51:11.3 &  230.5 & \nodata & \nodata & 19.2 & 15.21 & 14.00 & 10.65 & 1.21 & 3.35  \\ 
MALS0356+1900 & 03:56:33.46 & +19:00:34.6 & 1051.3 &  19.89  &  19.51  & 20.0 & 15.76 & 14.95 & 11.73 & 0.80 & 3.22  \\ 
MALS0356$-$0831 & 03:56:34.55 & $-$08:31:21.3 &  239.7 & \nodata & \nodata & 21.0 & 16.52 & 15.34 & 11.84 & 1.18 & 3.50  \\ 
MALS0455+1850 & 04:55:41.91 & +18:50:10.9 &  328.1 & \nodata & \nodata & 20.8 & 14.61 & 13.86 & 10.79 & 0.75 & 3.07  \\ 
MALS0510$-$1959 & 05:10:24.23 & $-$19:59:50.3 &  336.2 & \nodata & \nodata & 20.4 & 14.95 & 14.15 & 11.16 & 0.79 & 2.99  \\ 
MALS0512+1517 & 05:12:40.99 & +15:17:23.8 &  966.5 &  20.23  &  19.58  & 19.4 & 15.53 & 14.45 & 10.99 & 1.08 & 3.46  \\ 
MALS0516+0732 & 05:16:56.35 & +07:32:52.7 &  231.7 & \nodata & \nodata & 17.1 & 14.18 & 13.53 & 10.65 & 0.65 & 2.87  \\ 
MALS0519+1746 & 05:19:38.34 & +17:46:41.2 &  319.1 &  20.61  &  19.76  & 19.8 & 16.73 & 16.03 & 12.11 & 0.70 & 3.92  \\ 
MALS0529$-$1126 & 05:29:05.55 & $-$11:26:07.5 &  455.8 & \nodata & \nodata & 20.1 & 14.99 & 14.30 & 11.30 & 0.69 & 3.00  \\ 
MALS0600$-$0710 & 06:00:12.92 & $-$07:10:37.9 &  224.9 & \nodata & \nodata & 20.6 & 16.14 & 15.28 & 12.22 & 0.86 & 3.06  \\ 
MALS0730+1739 & 07:30:37.09 & +17:39:51.5 &  694.2 &  23.84  &  22.34  & 22.8 & 16.91 & 16.13 & 12.07 & 0.78 & 4.06  \\ 
MALS0731+1433 & 07:31:59.01 & +14:33:36.3 &  316.5 &  18.60  &  18.46  & 18.4 & 15.49 & 14.76 & 11.01 & 0.73 & 3.75  \\ 
MALS0824$-$1029 & 08:24:44.81 & $-$10:29:43.6 &  345.4 & \nodata & \nodata & 18.4 & 15.42 & 14.70 & 11.28 & 0.71 & 3.42  \\ 
MALS0836+0406 & 08:36:01.35 & +04:06:36.0 &  435.3 &  22.30  &  20.95  & 22.4 & 15.52 & 14.91 & 11.86 & 0.60 & 3.05  \\ 
MALS0845$-$1118 & 08:45:48.34 & $-$11:18:00.1 &  264.4 & \nodata & \nodata & \nodata & 16.18 & 15.29 & 12.03 & 0.89 & 3.25  \\ 
MALS0909$-$1637 & 09:09:10.66 & $-$16:37:53.8 &  340.1 & \nodata & \nodata & 20.1 & 15.87 & 15.20 & 11.84 & 0.67 & 3.36  \\ 
MALS0909$-$0500 & 09:09:16.99 & $-$05:00:53.2 &  457.2 & \nodata & \nodata & 21.1 & 15.68 & 15.00 & 11.75 & 0.68 & 3.25  \\ 
MALS0910$-$0526 & 09:10:51.01 & $-$05:26:26.8 &  337.9 & \nodata & \nodata & 17.8 & 14.57 & 13.61 & 10.46 & 0.96 & 3.16  \\ 
MALS0939$-$1731 & 09:39:19.21 & $-$17:31:35.4 &  263.3 & \nodata & \nodata & 20.4 & 15.19 & 14.07 & 10.67 & 1.12 & 3.40  \\ 
MALS0941$-$0321 & 09:41:46.01 & $-$03:21:21.7 &  346.2 & \nodata & \nodata & 21.2 & 16.59 & 15.77 & 11.94 & 0.82 & 3.83  \\ 
MALS0951$-$0601 & 09:51:38.34 & $-$06:01:38.0 &  378.8 & \nodata & \nodata & 20.5 & 16.27 & 15.46 & 11.95 & 0.82 & 3.50  \\ 
MALS0954+1252 & 09:54:14.75 & +12:52:22.6 &  355.7 & \nodata & \nodata & \nodata & 16.95 & 15.74 & 11.79 & 1.21 & 3.95  \\ 
MALS1007$-$1817 & 10:07:43.51 & $-$18:17:32.4 &  205.2 & \nodata & \nodata & 21.0 & 14.96 & 14.22 & 10.16 & 0.74 & 4.06  \\ 
MALS1008$-$0959 & 10:08:02.27 & $-$09:59:19.3 &  645.3 & \nodata & \nodata & 19.6 & 15.70 & 14.55 & 11.24 & 1.15 & 3.31  \\ 
MALS1009+1602 & 10:09:54.53 & +16:02:03.8 &  201.9 &  20.81  &  20.64  & 20.6 & 16.09 & 15.18 & 11.57 & 0.91 & 3.61  \\ 
MALS1024$-$0852 & 10:24:44.61 & $-$08:52:06.4 &  463.8 & \nodata & \nodata & 21.4 & 16.36 & 15.58 & 12.51 & 0.78 & 3.07  \\ 
MALS1033$-$1210 & 10:33:35.61 & $-$12:10:32.1 & 2065.1 & \nodata & \nodata & 20.9 & 16.59 & 15.80 & 11.70 & 0.80 & 4.10  \\ 
MALS1043+1641 & 10:43:47.34 & +16:41:06.9 &  259.1 &  22.15  &  20.52  & 21.3 & 15.82 & 15.19 & 11.74 & 0.63 & 3.45  \\ 
MALS1046+1535 & 10:46:08.12 & +15:35:36.6 &  430.9 & \nodata & \nodata & 22.1 & 16.83 & 16.03 & 11.88 & 0.80 & 4.15  \\ 
MALS1050$-$0318 & 10:50:24.43 & $-$03:18:08.8 &  243.5 &  18.74  &  18.70  & 18.6 & 15.25 & 14.06 & 10.77 & 1.19 & 3.29  \\ 
MALS1059$-$1118 & 10:59:55.36 & $-$11:18:18.7 &  314.5 & \nodata & \nodata & 20.6 & 16.60 & 15.71 & 12.18 & 0.89 & 3.53  \\ 
MALS1119$-$0527 & 11:19:17.36 & $-$05:27:07.9 & 1174.4 & \nodata & \nodata & 20.6 & 16.58 & 15.94 & 11.65 & 0.64 & 4.29  \\ 
MALS1124$-$1501 & 11:24:02.56 & $-$15:01:59.1 &  261.9 & \nodata & \nodata & 19.3 & 16.65 & 15.77 & 11.87 & 0.88 & 3.90  \\ 
MALS1130$-$1006 & 11:30:15.25 & $-$10:06:38.4 &  215.2 & \nodata & \nodata & \nodata & 16.26 & 15.50 & 12.00 & 0.76 & 3.51  \\ 
MALS1148+1404 & 11:48:25.45 & +14:04:49.7 &  322.0 &  22.89  &  21.66  & 22.6 & 16.27 & 15.61 & 11.79 & 0.67 & 3.82  \\ 
MALS1150+1317 & 11:50:43.70 & +13:17:36.1 &  254.4 &  20.12  &  19.88  & 19.7 & 16.31 & 15.24 & 12.02 & 1.07 & 3.22  \\ 
MALS1206$-$0714 & 12:06:32.23 & $-$07:14:52.6 &  698.8 & \nodata & \nodata & 20.5 & 16.03 & 15.41 & 11.68 & 0.62 & 3.73  \\ 
MALS1215$-$0628 & 12:15:14.42 & $-$06:28:03.5 &  360.4 & \nodata & \nodata & 19.8 & 15.78 & 15.01 & 11.76 & 0.76 & 3.26  \\ 
MALS1218$-$0631 & 12:18:36.18 & $-$06:31:15.8 &  409.1 & \nodata & \nodata & 18.9 & 14.35 & 13.50 & 10.41 & 0.85 & 3.09  \\ 
MALS1219$-$1809 & 12:19:05.45 & $-$18:09:11.2 &  300.3 & \nodata & \nodata & 19.5 & 14.80 & 14.15 & 11.26 & 0.65 & 2.89  \\ 
MALS1221$-$1237 & 12:21:23.49 & $-$12:37:24.1 &  343.7 & \nodata & \nodata & 20.1 & 15.67 & 14.79 & 11.77 & 0.88 & 3.02  \\ 
MALS1225$-$1144 & 12:25:24.48 & $-$11:44:31.2 &  479.5 & \nodata & \nodata & 20.9 & 14.90 & 14.26 & 11.24 & 0.64 & 3.03  \\ 
MALS1231$-$1236 & 12:31:50.30 & $-$12:36:37.5 &  276.0 & \nodata & \nodata & 19.2 & 15.00 & 13.86 & 10.42 & 1.14 & 3.44  \\ 
MALS1318$-$1441 & 13:18:56.01 & $-$14:41:55.0 &  283.0 & \nodata & \nodata & 19.8 & 15.14 & 14.08 & 10.67 & 1.05 & 3.42  \\ 
MALS1343$-$0834 & 13:43:09.26 & $-$08:34:57.5 &  241.5 & \nodata & \nodata & 19.8 & 15.93 & 15.09 & 11.88 & 0.84 & 3.21  \\ 
MALS1351$-$1019 & 13:51:31.98 & $-$10:19:32.9 &  726.1 & \nodata & \nodata & 19.3 & 15.02 & 14.12 & 10.74 & 0.90 & 3.38  \\ 
MALS1456+0456 & 14:56:25.83 & +04:56:45.2 &  287.9 &  20.63  &  20.29  & 20.5 & 16.21 & 15.50 & 12.62 & 0.71 & 2.88  \\ 
MALS1623+1239 & 16:23:03.03 & +12:39:58.4 &  523.3 &  18.60  &  18.23  & 18.1 & 16.16 & 15.50 & 12.16 & 0.66 & 3.34  \\ 
MALS1625$-$0416 & 16:25:39.34 & $-$04:16:16.3 &  257.2 &  23.05  &  22.69  & 19.1 & 16.39 & 15.75 & 12.47 & 0.64 & 3.28  \\ 
MALS1639+1144 & 16:39:06.46 & +11:44:09.2 &  341.9 &  20.72  &  19.62  & 19.6 & 16.17 & 15.13 & 11.48 & 1.03 & 3.66  \\ 
MALS1645$-$0432 & 16:45:52.10 & $-$04:32:53.3 &  423.3 & \nodata & \nodata & 20.7 & 16.78 & 15.94 & 12.32 & 0.84 & 3.62  \\ 
MALS1649+0626 & 16:49:50.51 & +06:26:53.3 &  389.2 & \nodata & \nodata & 18.6 & 14.75 & 13.64 & 10.40 & 1.11 & 3.23  \\ 
MALS1650$-$1248 & 16:50:38.03 & $-$12:48:54.5 &  275.5 & \nodata & \nodata & 20.8 & 16.23 & 15.07 & 11.27 & 1.17 & 3.79  \\ 
MALS1653$-$0102 & 16:53:57.67 & $-$01:02:14.2 &  317.4 &  21.27  &  20.47  & 20.4 & 15.71 & 14.80 & 11.67 & 0.91 & 3.13  \\ 
MALS1716$-$0613 & 17:16:27.09 & $-$06:13:56.5 &  229.9 & \nodata & \nodata & 19.4 & 14.03 & 12.82 & 9.49 & 1.21 & 3.33  \\ 
MALS1721+1626 & 17:21:05.79 & +16:26:49.1 &  507.9 & \nodata & \nodata & 19.3 & 16.11 & 15.31 & 11.80 & 0.79 & 3.51  \\ 
MALS1722+1652 & 17:22:39.45 & +16:52:08.7 &  257.1 & \nodata & \nodata & 19.9 & 16.29 & 15.50 & 12.31 & 0.80 & 3.19  \\ 
MALS1724+0326 & 17:24:23.07 & +03:26:32.5 &  217.8 & \nodata & \nodata & 18.7 & 16.31 & 15.23 & 11.61 & 1.08 & 3.62  \\ 
MALS1725+0622 & 17:25:07.45 & +06:22:42.1 &  213.4 & \nodata & \nodata & 18.6 & 15.40 & 14.49 & 11.36 & 0.91 & 3.14  \\ 
MALS2034$-$0523 & 20:34:25.65 & $-$05:23:32.2 &  419.7 & \nodata & \nodata & 18.6 & 15.58 & 14.61 & 11.18 & 0.97 & 3.43  \\ 
MALS2054$-$0932 & 20:54:56.08 & $-$09:32:40.8 &  313.6 & \nodata & \nodata & 20.8 & 16.42 & 15.44 & 11.74 & 0.98 & 3.70  \\ 
MALS2059$-$1440 & 20:59:59.61 & $-$14:40:43.1 & 1018.1 & \nodata & \nodata & 21.2 & 14.63 & 13.63 & 10.45 & 0.99 & 3.18  \\ 
MALS2120+1327 & 21:20:42.48 & +13:27:24.2 &  401.5 &  20.02  &  19.56  & 19.5 & 14.68 & 13.91 & 10.56 & 0.77 & 3.35  \\ 
MALS2139+1718 & 21:39:37.03 & +17:18:26.9 &  219.4 &  19.26  &  19.06  & 18.9 & 15.43 & 14.23 & 10.88 & 1.21 & 3.34  \\ 
MALS2158+0925 & 21:58:00.88 & +09:25:46.4 &  437.7 &  21.01  &  19.78  & 20.5 & 15.19 & 14.33 & 9.91 & 0.87 & 4.41  \\ 
MALS2201+0312 & 22:01:27.50 & +03:12:15.6 &  300.5 &  23.58  &  22.75  & 20.3 & 16.31 & 15.61 & 11.78 & 0.70 & 3.83  \\ 
MALS2220+1307 & 22:20:04.97 & +13:07:12.1 &  812.0 &  20.46  &  20.16  & 20.4 & 14.94 & 14.19 & 11.11 & 0.74 & 3.08  \\ 
MALS2223+1213 & 22:23:59.11 & +12:13:38.9 &  246.4 &  21.70  &  20.40  & 21.5 & 15.05 & 13.99 & 10.76 & 1.06 & 3.23  \\ 
MALS2224+1304 & 22:24:20.03 & +13:04:50.2 &  294.5 &  20.91  &  20.62  & 20.8 & 16.15 & 15.13 & 11.67 & 1.02 & 3.46  \\ 
MALS2230+0348 & 22:30:50.19 & +03:48:36.8 &  242.8 &  22.48  &  21.79  & 22.9 & 16.53 & 15.92 & 12.15 & 0.61 & 3.77  \\ 
MALS2238$-$1344 & 22:38:26.48 & $-$13:44:22.6 &  241.2 & \nodata & \nodata & 20.0 & 14.78 & 13.97 & 10.78 & 0.81 & 3.19  \\ 
MALS2243+1814 & 22:43:54.81 & +18:14:45.9 & 1004.4 &  20.98  &  20.23  & 20.8 & 15.94 & 14.48 & 10.34 & 1.47 & 4.13  \\ 
MALS2247+1211 & 22:47:05.52 & +12:11:51.4 &  223.7 & \nodata & \nodata & 18.2 & 17.27 & 16.66 & 12.25 & 0.61 & 4.41  \\ 
MALS2258$-$0958 & 22:58:13.47 & $-$09:58:17.2 &  225.7 &  25.59  &  26.20  & \nodata & 17.37 & 16.52 & 12.19 & 0.84 & 4.34  \\ 
MALS2300+1940 & 23:00:36.41 & +19:40:02.9 &  210.4 &  19.54  &  19.48  & 19.6 & 15.94 & 15.22 & 12.19 & 0.72 & 3.03  \\ 
MALS2300+0337 & 23:00:40.87 & +03:37:10.3 &  509.0 &  19.68  &  19.60  & 19.9 & 15.54 & 14.45 & 10.89 & 1.09 & 3.56  \\ 
MALS2308$-$1149 & 23:08:05.25 & $-$11:49:45.7 &  256.8 & \nodata & \nodata & 20.0 & 16.66 & 15.90 & 12.07 & 0.77 & 3.82  \\ 
MALS2310+1114 & 23:10:02.89 & +11:14:03.6 &  248.1 &  19.95  &  19.82  & 20.0 & 16.55 & 15.39 & 12.12 & 1.16 & 3.27  \\ 
MALS2316+0429 & 23:16:34.61 & +04:29:40.2 &  214.0 &  19.09  &  18.97  & 19.2 & 15.72 & 14.46 & 11.15 & 1.26 & 3.31  \\ 
MALS2332$-$1423 & 23:32:31.61 & $-$14:23:18.8 &  234.1 & \nodata & \nodata & 20.5 & 14.79 & 14.10 & 11.06 & 0.68 & 3.04  \\ 
MALS2338$-$1218 & 23:38:08.04 & $-$12:18:51.4 &  450.9 & \nodata & \nodata & 19.6 & 16.01 & 15.04 & 11.81 & 0.97 & 3.23  \\ 
MALS2340+0959 & 23:40:07.26 & +09:59:59.0 &  205.0 &  21.89  &  21.63  & 22.6 & 15.42 & 14.81 & 11.95 & 0.62 & 2.86  \\ 
MALS2359+1924 & 23:59:14.02 & +19:24:20.6 &  284.1 &  21.45  &  20.67  & 20.8 & 14.46 & 13.58 & 10.18 & 0.88 & 3.40  \\ 
\enddata

\tablecomments{
(1) Target identifier; (2) Right Ascension (epoch J2000); (3) Declination (epoch J2000);\\
(4) Flux at $1.4$~GHz in mJy; (5) SDSS $g$-band AB magnitude; (6) SDSS $r$-band AB magnitude;\\
(7) Synthetic SDSS $r$-band magnitude calculated from the observed spectrum (typical uncertainty of $\pm0.2$~mag);\\
(8)--(10) {\it WISE} bands 1, 2 and 3 at 3.4, 4.6, and 12~$\mu$m in Vega magnitudes;\\
(11)--(12) {\it WISE} colors in Vega magnitudes.
}

\end{deluxetable*}

\end{document}